\documentstyle[multicol,aps,epsfig]{revtex} 

\tighten       

\begin{document}
\draft 

\title{Motion of Wavefronts in Semiconductor Superlattices}
\author{ A. Carpio $^1$, L. L. Bonilla $^2$ and G. Dell'Acqua
$^2$}
\address{$^1$ Departamento de Matem\'atica
Aplicada, Universidad Complutense, Madrid 28040, Spain\\
$^2$ Departamento de Matem\'{a}ticas, Escuela
Polit\'ecnica Superior,  Universidad Carlos III de Madrid,\\
Avenida de la Universidad 30, 28911 Legan{\'e}s, Spain\\
Also: Unidad Asociada al Instituto de Ciencia de Materiales
(CSIC)}
\date{ \today  }
\maketitle

\begin{abstract}
An analysis of wavefront motion in weakly coupled doped
semiconductor superlattices is presented. If a dimensionless
doping is sufficiently large, the superlattice behaves as a
discrete system presenting front propagation failure and the
wavefronts can be described near the threshold currents $J_i$
($i=1,2$) at which they depin and move. The wavefront velocity
scales with current as $|J-J_i|^{{1\over 2}}$. If the
dimensionless doping is low enough, the superlattice behaves as
a continuum system and wavefronts are essentially shock waves
whose velocity obeys an equal area rule.
\end{abstract}

\begin{multicols}{2}
\narrowtext

\setcounter{equation}{0}
\section{Introduction}
\label{sec:intro}
Semiconductor superlattices (SL) are unique nonlinear systems.
Experimental evidence shows features of spatially discrete
systems (multistability of roughly as many stationary states as
SL periods due to formation of electric field domains
\cite{gra91}) in certain regimes of nonlinear charge transport
in SL. In other regimes (self-sustained periodic and chaotic
oscillations of the current \cite{kas95}), SL behavior is more
typical of continuous systems. These different properties may be
described by discrete \cite{bon94} or continuous balance
equation models \cite{but78}. The continuum limit of discrete
SL models has been used to understand important aspects of
self-oscillations \cite{bon97}. On the other hand, discrete
models of SL share common characteristics with spatially
discrete systems in other fields: most importantly, front
propagation failure and front depinning \cite{car00}.

Discrete systems describe physical reality in many different
fields: propagation of nerve impulses along mielinated fibers
\cite{kee87,kee98}, pulse propagation through cardiac cells
\cite{kee98}, calcium release waves in living cells
\cite{bug97}, sliding of charge density waves \cite{cdw},
superconductor Josephson array junctions \cite{jj}, arrays of
coupled diode resonators \cite{diode}, motion of dislocations
in crystals \cite{nab67} and atoms adsorbed on a periodic
substrate \cite{cha95}. A distinctive feature of discrete
systems (not shared by continuous ones) is the phenomenon of
wavefront pinning: for values of a control parameter in a
certain interval, wavefronts joining two different
constant states fail to propagate \cite{kee98}.
When the control parameter surpasses a threshold,
the wavefront depins and starts moving
\cite{car00,kee87,cdw,nab67}. The existence of
such thresholds is thought to be an intrinsecally
discrete fact, which is lost in continuum
aproximations. The characterization of propagation
failure and front depinning in discrete systems
is thus an important problem, which is still
poorly understood despite the numerous inroads
made in the literature \cite{kee87,bug97,cdw,nab67}.

Recently, several of us have proposed a theory of front
depinning and propagation failure in discrete
reaction-diffusion (RD) systems \cite{CB}. This theory will be
extended here to describe motion of wavefronts near depinning
thresholds in discrete drift-diffusion (DDD) models of SL
\cite{bon00}. There are important differences between RD and SL
systems. Let us consider a RD system depending on a parameter
$A$, measuring how close we are to the continuum limit ($A=0$),
and an external parameter (`the field' $F$) such that wavefronts
move for all values of $F\neq 0$ in the continuum limit. In a
generic RD system, there is always a pinning interval
(wavefronts are stationary for $F$ in the pinning interval)
about $F=0$ no matter how close to the continuum limit we are.
As we approach the continuum limit, the pinning interval
shrinks to zero exponentially fast as $A$ approaches zero: we
then need exponential asymptotics (asymptotics beyond all
orders) to describe what happens. In a SL, there are important
differences: (i) the pinning interval disappears before we
reach the continuum limit, and (ii) the continuum limit is
described by hyperbolic equations and shock waves (no
exponential asymptotics is needed).

 Let us consider an infinitely long, sufficiently doped SL under
constant current conditions. The current density plays the
role of external field in the SL models and a dimensionless
doping $\nu$ plays the same role as the parameter $A$: the
continuum limit is reached as $\nu\to 0$. For large values of
$\nu$, we are in the strongly discrete limit and wavefronts
are stationary (pinned) for currents on a certain pinning
interval. If the current is smaller than the lowest limit of
the interval, $J_{1}(\nu)$, the wavefront moves downstream,
following the electron flow. For currents larger than the upper
limit of the interval, $J_{2}(\nu)$, wavefronts move upstream,
against the electron flow \cite{car00}. As the doping decreases,
first $J_{2}$ and then $J_{1}$ (for even lower doping)
disappear. This means that fronts are either stationary or move
downstream below a first critical doping and they always move
downstream below a second critical doping. Since the continuum
limit corresponds to vanishing dimensionless doping, stationary
fronts and fronts moving upstream are features of discrete SL,
which are lost in the continuum limit.

\begin{figure}
\includegraphics[width=8cm]{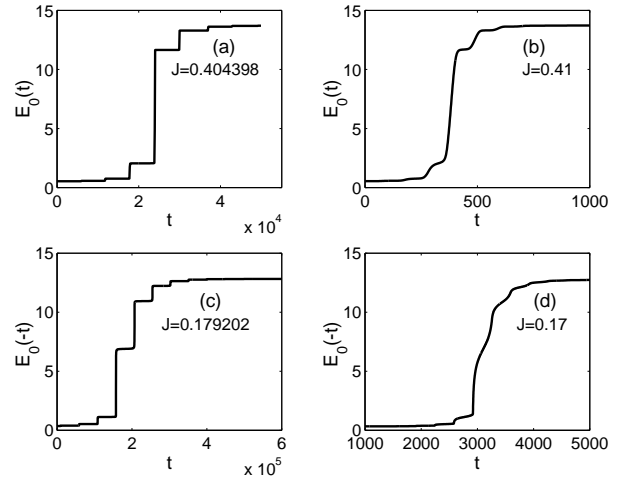}
\vspace{0.5 cm}
\caption{Sharpening of wavefront profiles as the dimensionless
current $J$ approaches its critical values for $\nu=3$. (a) $J
\approx J_2$, (b) $J>J_2$, (c) $J\approx J_1$, (d) $J<J_1$. }
\label{fig1}
\end{figure}

In contrast with the precipituous jumps and falls of charge and
electric field in individual wells during wavefront motion, we
will show that, for currents outside $(J_{1},J_{2})$,
wavefronts are described by continuous profiles moving at a
constant velocity. These profiles become sharper as the
currents approach the threshold values $J_{i}$; see Fig.\
\ref{fig1}. Exactly at these values, wavefront profiles cease
to be continuous: a number of jump discontinuities open up,
which results in propagation failure. We describe propagation
failure by analyzing the behavior of several {\em active} wells
which govern front motion. Our theory yields the front profile
near threshold currents and a universal scaling of the front
velocity for sufficiently large doping. Its performance worsens
as the doping decreases and we approach the continuum limit. We
will complement our approach by obtaining the velocity of
wavefronts and their shape in this limit. Previously, these
front profiles and velocities in the continuum limit were known
only for models having no discrete diffusion \cite{bon95},
which is correct only for high enough voltage bias. For bias on
the first plateau of the current--voltage characteristics,
diffusivity cannot be ignored. Given that characterizing
wavefronts at constant current is the key to analyze
self-oscillations at constant voltage, the results in this
paper could be useful to describe self-oscillations at low
biases. This could be done in two limits: near the continuum
limit $\nu\ll 1$, the results we have obtained for shock waves
and monopoles can be used to characterize self-oscillations,
either mediated by monopoles \cite{bon97} or by dipoles
\cite{hig92},\cite{gunn}. For larger $\nu$, our present theory
paves the way to understanding self-oscillations in DDD or more
general SL models.

The rest of the paper is as follows. We describe front
propagation failure and front depinning for DDD SL models in
Section \ref{sec:2}. The continuum limit of the model is
analyzed in Section \ref{sec:3}. The resulting equations are
those of the Kroemer model of the Gunn effect with zero
diffusivity. Wavefronts are shock wave solutions of this model
with a rigid tail region either to the right or left of the
shock. The shock velocity must be obtained from the discrete
model, which results in an equal-area rule rather different
from that of the Gunn effect. Corrections to this rule are
important for describing the self-oscillations and will be given
here. Section \ref{sec:conclusions} contains our conclusions.
Lastly, the Appendices are devoted to different technical
matters.

\section{Propagation failure and front depinning}
\label{sec:2}
In weakly coupled SL, (inter and intra-subband) scattering
times are much shorter than well escape times, which in turn
are much shorter than macroscopic times (period of
self-oscillations). Then the dominant mechanism of vertical
transport is sequential tunneling, only the first subband of
each well is appreciably occupied and the tunneling current
across barriers is stationary \cite{bon94}. Nonlinear phenomena
seen in experiments can be described by discrete balance
equations.

We count the barrier separating the injecting contact from the
first well of the SL as the zeroth barrier. Barriers and wells
have widths $d$ and $w$, respectively, so that $l=d+w$ is the
SL period. The $i$th SL period starts just before the $i$th
barrier and ends just before the $(i+1)$st barrier. With this
convention, the dependent variables of our model are {\em minus}
the electric field averaged over the ith period, $F_i$, and the
two-dimensional electron density at the $i$th well
(concentrated in a plane normal to the growth direction,
located at the end of the $i$th well), $n_i$. These variables
obey the Poisson and charge continuity equations:
\begin{eqnarray}
F_{i}-F_{i-1} = {e\over\varepsilon}\, (n_{i}-N_{D}^{w}) .
\label{m1}\\
{dn_{i}\over dt} = J_{i-1\to i} - J_{i\to i+1}. \label{m2}
\end{eqnarray}
Here $N_D^w$, $\varepsilon$, $e$ and $e J_{i\to i+1}$ are the 2D
doping density at the $i$th well, the average permittivity of
the SL, minus the electron charge and the tunneling current
density across the $i$th barrier, respectively. We can
differentiate (\ref{m1}) with respect to time and eliminate
$n_i$ by using (\ref{m2}). The result can be written as a form
of Amp\`ere's law for the balance of current
\begin{eqnarray}
{\varepsilon\over e}\, {dF_{i}\over dt} + J_{i\to i+1} =
J(t)\,.\label{ampere}
\end{eqnarray}
Here $e J(t)$ is the total current density through the SL,
equal for all SL periods, and $\varepsilon\, dF_i/dt$ is the
displacement current at the $i$th SL period.

The tunneling current density $e J_{i\to i+1}$ is related to
electric fields and electron densities by a {\em constitutive
relation}, which should be derived from first principles; see
Wacker in Ref.\ \onlinecite{bon94} and Ref.\
\onlinecite{bon00}. At sufficiently low or high temperatures,
$J_{i\to i+1}$ has the following drift-diffusion form
\cite{bon00},
\begin{eqnarray}
J_{i\to i+1}= {n_{i}v(F_{i})\over l} - D(F_i)\,
{n_{i+1}-n_{i}\over l^{2}} , \label{m3}
\end{eqnarray}
where the {\em drift} velocity is an odd function of the field,
$v(-F)=-v(F)$, and the {\em diffusion coefficient} satisfies
the relation
\begin{eqnarray}
D(-F) = v(F)\, l + D(F) . \label{m4}
\end{eqnarray}
Typical forms of these coefficients (in nondimensional form;
see below) are shown in Figure \ref{fig2}. To compare our
theoretical results with numerical solutions of the model, it
is better to use analytical approximations of these functions.
Ours are given in Appendix \ref{app1}. The Amp\`ere's law
corresponding to the current (\ref{m3}) is
\begin{eqnarray}
{\varepsilon\over e}\, {dF_{i}\over dt} + {n_{i}v(F_{i})\over
l} - D(F_i)\, {n_{i+1}-n_{i}\over l^{2}}= J(t)\,.
\label{m5}
\end{eqnarray}
Equations (\ref{m1}) and (\ref{m5}) form the DDD model
\cite{car00,bon00}, studied here. Notice that for high fields
(in practice for all plateaus except the first one), $D=0$, and
we have a simpler discrete drift model with $J_{i\to i+1}=
n_{i}v(F_{i})/l$ \cite{bon94}.

To analyze the discrete drift-diffusion model, it is convenient
to render  all equations dimensionless. Let $v(F)$ reach its
first positive maximum at $(F_M,v_M)$. We adopt $F_M$,
$N_D^w$, $v_M$, $v_M\, l$, $eN_D^w v_M/l$ and $\varepsilon F_M
l/(e N_D^w v_M)$ as the  units of $F_i$, $n_i$, $v(F)$, $D(F)$,
$eJ$ and $t$, respectively.  For the first plateau of the 9/4
SL of Ref.\ \onlinecite{kas95}, we  find $F_M = 6.92$ kV/cm,
$N_D^w = 1.5\times 10^{11}$ cm$^{-2}$, $v_M =  156$ cm/s,
$v_M\, (d+w) = 2.03\times 10^{-4}$ cm$^2 /$s and $eN_D^w
v_M/(d+w) = 2.88$ A/cm$^2$. The units of current and time are
0.326 mA  and 2.76 ns, respectively. Then (\ref{m1}) and
(\ref{m5}) become
\begin{eqnarray}
{dE_{i}\over dt} + v(E_i)\, n_i - D(E_i)\, (n_{i+1}-n_i) = J,
\label{m6}\\
E_i - E_{i-1} = \nu\, (n_i - 1) , \label{m7}
\end{eqnarray}
or, equivalently,
\begin{eqnarray}
{dE_{i}\over dt} + v(E_i)\, {E_{i}- E_{i-1}\over\nu} -
D(E_i)\, {E_{i+1}+E_{i-1}-2E_{i}\over\nu}\nonumber\\
= J - v(E_i). \label{m8}
\end{eqnarray}
Here we have used the same symbols for dimensional and
dimensionless quantities except for the electric field ($F$
dimensional, $E$ dimensionless). $\nu=e N_D^w/(\varepsilon\,
F_M)$ is the dimensionless doping parameter, which is about 3
for the first plateau of the 9/4 SL.

\subsection{Phase diagram for wavefronts on an infinite
superlattice}
Here we shall consider an infinite SL under constant current
bias $J$. Clearly, there are two stable spatially homogeneous
stationary solutions, namely $E^{(1)}(J)$ and $E^{(3)}(J)$,
where $v(E^{(k)})=J$, $E^{(1)}(J)<E^{(2)}(J)<E^{(3)}(J)$. We
are interested in nonuniform front states of the DDD model which
satisfy $E_i\to E^{(1)}(J)$ as $i\to -\infty$ and $E_i \to
E^{(3)}(J)$ as $i\to \infty$. These states are either
stationary or time-dependent. In the second case, they are
wavefronts moving with constant velocity $c= c(J,\nu)$, such
that $E_i(t)= E(i-ct)$, $i=0,\pm 1,\ldots$ $E(\tau)$ is a smooth
profile which solves the following nonlinear eigenvalue
problem for $c$ (measured in wells traversed per unit time) and
$E(\tau)$:
\begin{eqnarray}
c\, {dE\over d\tau} = v(E) - J + v(E)\, {E- E(\tau-1)\over\nu}
\nonumber\\
- D(E)\, {E(\tau+1)+E(\tau-1)-2E\over\nu}\,, \label{dde}\\
E(-\infty)= E^{(1)}(J),\quad\quad E(\infty)= E^{(3)}(J) .
\label{bc}
\end{eqnarray}
By using a comparison principleÊ\cite{comparison}, we can prove
the existence of stationary fronts rigorously \cite{car00}. 
Outside the interval of current values in which there are
stationary fronts, we can only prove that there are fronts
moving to the right or the left \cite{car00}. By using the
comparison principle, it is possible to show that moving and
stationary fronts cannot exist simultaneously at the same value
of the current \cite{car99}. This result does not hold if the
spatially discrete differential equation is second order in
time: moving and stationary fronts have been reported to coexist
in one such case \cite{fla99}. 

Proofs that moving fronts are traveling wavefronts have been
given for different spatially discrete models without convection
\cite{existence,car99}. In the case of the DDD SL model, we
rely on numerical evidence: Notice that all noticeable steps in
Fig.\ \ref{fig1}, particularly (a) and (c), are of the same
length. Fig.\ \ref{fig3} shows the wavefront at three different
times, and it clearly demonstrates that the front is a
traveling wave moving with constant velocity as a whole. Our
asymptotic construction of the wavefronts near the critical
currents $J_1$ and $J_2$, which we explain below, exploits
their traveling wave nature.

Solving numerically (\ref{m8}), it can be shown that, after a
short transient, a variety of initial conditions such that
$E_i\to E^{(1)}(J)$ as $i\to -\infty$ and $E_i \to E^{(3)}(J)$
as $i\to\infty$ evolve towards either a stationary or moving
monopole. Figure 3 of Ref.\ \onlinecite{car00} is a phase
diagram showing the regions in the plane $(J,\nu)$ where
different fronts are stable. There are two important values of
$\nu$, $\nu_1<\nu_2$, such that:
\begin{itemize}
\item For $0<\nu<\nu_1$ and each fixed $J\in (v_m,1)$, only
traveling monopole fronts moving downstream (to the right) were
observed. For $\nu>\nu_1$, stationary monopoles were found.
\item For $\nu_1<\nu<\nu_2$, traveling fronts moving downstream
exist only if $J\in (v_{m},J_1(\nu))$, where $J_1(\nu)\in
(v_m,1)$ is a critical value of the current. $J_1(\nu)$ is a
monotone decreasing function such that $J_1\to v_m$ as $\nu\to
\infty$; see Fig.\ \ref{fig4}. If $J\in (J_1(\nu),1)$, the
stable solutions are steady fronts (stationary monopoles).
\item New solutions are observed for $\nu>\nu_2$. As before,
there are traveling fronts moving downstream if $J\in (v_{m},
J_1(\nu))$, and stationary monopoles if $J \in (J_1(\nu),
J_2(\nu))$, $J_2(\nu)<1$ is a new  critical current; see Fig.\
\ref{fig4}. The function $J_2(\nu)$ starts at $J_2(\nu_2)=1$,
decreases to a single minimum value, and then it increases
towards 1 as $\nu\to\infty$. For $J_2(\nu)<J<1$, the stable
solutions of (\ref{m9}) and (\ref{m10}) are monopoles traveling
upstream (to the left). As $\nu$ increases, $J_1(\nu)$ and
$J_2(\nu)$ approach $v_m$ and 1, respectively. Thus stationary
solutions are found for most values of $J$ if $\nu$ is large
enough.
\end{itemize}

Wavefront velocity as a function of current has been depicted in
Figure \ref{fig5} for $\nu = 3$, which corresponds to the first
plateau of the 9/4 SL. For larger $\nu$, the interval of $J$
for which stationary solutions exist becomes wider again,
trying to span the whole interval $(v_m,1)$ as $\nu\to \infty$.
For very large $\nu$, the velocities of downstream and upstream
moving monopoles become extremely small in absolute value.

\subsection{Pinning of wavefronts with a single active well}
At the critical currents, $J_1(\nu)$ and $J_2(\nu)$, wavefronts
moving downstream (to the right, following the electron flow,
$c>0$) for smaller $J$ or upstream (to the left, against the
electron flow, $c<0$) for larger $J$ fail to propagate. What
happens is that the wavefront field profile $E(\tau)$ becomes
sharper as $J$ approaches the critical currents. Exactly at
$J_k$, $k=1,2$, gaps open up in the wavefront profile which
therefore loses continuity. The resulting field profile is a
stationary front $E_i= E_i(J,\nu)$: the wavefront is pinned for
$J_1<J<J_2$. The depinning transition (from stationary fronts
to moving wavefronts) is technically speaking a global
saddle-node bifurcation. We shall study it first in the
simplest case of large dimensionless doping $\nu$, and then
indicate what happens in the general case.

For sufficiently large doping and $J$ close to critical, the
moving front is led by the behavior of a single well, which we
will call the {\em active well}. If we examine the shape of a
stationary front near the critical current, we observe that all
wells are close either to $E^{(1)}(J)$ or $E^{(3)}(J)$ except
for one well which drifts slowly and eventually jumps: the
active well. Let us call $E_0$ the electric field at the active
well. Since all wells in the front perform the same motion, we
can reconstruct the profile $E(i-ct)$ from the time evolution
of $E_0(t)= E(-ct)$. Before the active well jumps, $E_i\approx
E^{(1)}(J)$ for $i<0$ and $E_i\approx E^{(3)}(J)$ for $i>0$.
Thus Eq.\ (\ref{m8}) becomes
\begin{eqnarray}
{dE_{0}\over dt} \approx J - v(E_0) -v(E_0)\, {E_{0}-
E^{(1)}\over\nu} \nonumber\\
+ D(E_0)\, {E^{(1)} + E^{(3)} -2 E_{0}\over\nu}\,. \label{m9}
\end{eqnarray}
This equation has three stationary solutions for $J_1<J<J_2$,
two stable and one unstable, and only one stable stationary
solution otherwise. At the critical currents, two of these
solutions coalesce forming a saddle-node. At low values
of the current, the two coalescing solutions form a double zero
corresponding to a maximum of the right side of (\ref{m9}).
For high currents, the two coalescing solutions form a double
zero corresponding to a minimum of the right side of (\ref{m9}).
The critical currents are such that the expansion of the right
hand side of (\ref{m9}) about the two coalescing stationary
solutions,
\begin{eqnarray}
J - v(E_0)-v(E_0)\, {E_{0}- E^{(1)}\over\nu} \nonumber\\
+ D(E_0)\, {E^{(1)} + E^{(3)} -2 E_{0}\over\nu} = 0, \label{m10}
\end{eqnarray}
has zero linear term,
\begin{eqnarray}
D'_{0} (E^{(1)} + E^{(3)} -2 E_{0})- 2 D_{0} \nonumber\\
- v'_{0} (E_{0}-E^{(1)}) - v_{0} -\nu\, v'_{0}
= 0. \label{m11}
\end{eqnarray}
Here $D'_0$ means $D'(E_0)$, etc. Equations (\ref{m10}) and
(\ref{m11}) yield approximations to $E_0$ and the critical
current $J_c$ (which is either $J_1$ or $J_2$). The results are
depicted in Figs.\ \ref{fig5} and \ref{fig6}, and show excellent
agreement with those of numerical solutions of the model for
$\nu>2$. Our approximation performs less well for smaller
$\nu$, which indicates that more active wells are needed to
improve it.

Let us now construct the profile of the traveling wavefronts
after depinning, for $J$ sligthly below $J_1$ or slightly above
$J_2$. Up to terms of order $|J-J_c|$, Equation (\ref{m9})
becomes
\begin{eqnarray}
{d\varphi\over dt} \approx \alpha\, (J - J_c) +
\beta\,\varphi^{2}\,,
\label{m12}
\end{eqnarray}
for $E_0(t) = E_0(J_c) + \varphi(t)$, as $J\to J_c$. $E_0(J_c)$
is the stationary solution of (\ref{m9}) at $J=J_c$. The
coefficients $\alpha$ and $\beta$ are given by
\begin{eqnarray}
\alpha = 1 + {v_{0}+D_{0}\over\nu\, v'_{1}} + {D_{0}\over \nu\,
v'_{3}}\,,\label{m13}\\
2\nu\beta = D''_0
(E^{(1)} + E^{(3)} -2 E_{0})- 4 D'_0 - 2 v'_{0}
\nonumber\\
+ v''_0 (E^{(1)} - E_{0}- 2\nu)  . \label{m14}
\end{eqnarray}
$\beta$ is negative if $J_c=J_1$ and positive if $J_c=J_2$. Eq.\
(\ref{m12}) has the (outer) solution
\begin{equation}
\varphi(t)\sim (-1)^k \sqrt{{(\alpha\, (J-J_{k})\over \beta}}\,
\tan\left(\sqrt{\alpha\beta\, (J-J_{k})}\, (t-t_0)\right)
\label{outer}
\end{equation}
($k=1,2$), for $J$ such that sign$(J-J_k)=$sign$\beta$.
(\ref{outer}) is very small most of the time, but it blows up
when the argument of the tangent function approaches $\pm
\pi/2$. Thus the outer approximation holds over a time interval
$(t-t_0)\sim \pi/\sqrt{\alpha\beta\, (J-J_{k})}$. The reciprocal
of this time interval yields an approximation for the wavefront
velocity,
\begin{equation}
|c(J,\nu)|\sim {\sqrt{\alpha\beta\, (J-J_{k})}\over\pi}\,.
\label{c}
\end{equation}
In Figs.\ \ref{fig5} and \ref{fig6}, we compare this
approximation with the numerically computed velocity for
$\nu=3$ and $\nu=20$, respectively.

When the solution begins to blow up, the outer solution
(\ref{outer}) is no longer a good approximation, for $E_0(t)$
departs from the stationary value $E_0(J_c)$. We must go back
to (\ref{m9}) and obtain an inner approximation to this
equation. As $J$ is close to $J_c$ and $E_0(t)-E_0(J_c)$ is of
order 1, we solve numerically (\ref{m12}) at $J=J_c$ with the
matching condition that $E_0(t)-E_0(J_c)\sim (-1)^k 2/[\pi
\sqrt{\beta/[\alpha\, (J-J_c)]} - 2|\beta|\, (t-t_0)]$, as
$(t-t_0)\to -\infty$. This inner solution describes the jump of
$E_0$ to values close to $E^{(1)}$ if $J_c=J_1$, or to values
close to $E^{(3)}$ if $J_c=J_2$. During this jump, the motion of
$E_0$ forces the other points to move. Thus, for $J_c=J_1$,
$E_{1}(t)$ can be calculated by using the inner solution in
(\ref{m8}) for $E_0$, with $J= J_c$ and $E_{2}\approx E^{(3)}$.
Similarly, for $J_c=J_2$, $E_{-1}(t)$ can be calculated by
using the inner solution in (\ref{m8}) for $E_0$, with $J= J_c$
and $E_{-2}\approx E^{(1)}$. A composite expansion \cite{bon87}
constructed with these inner and outer solutions is compared to
the numerical solution of the model in Fig.\ \ref{fig7}. Notice
that we have reconstructed the traveling wave profiles
$E(i-ct)$ from the identity $E_0(t)=E(-ct)$.

\subsection{Pinning of wavefronts with several active wells}
The previous asymptotic description of the depinning transition
deteriorates as $\nu$ decreases. What happens is that we need
more than one active well to approximate wavefront motion.
Depinning is then described by a reduced system of more than
one degree of freedom corresponding to active wells. There is a
saddle-node bifurcation in this reduced system whose normal
form is again (\ref{m12}) with different coefficients. The jump
of the active wells after blow up is found by solving the
reduced system with a matching condition. How do we determine
the optimal number of active wells? For a given $\nu$, the
eigenvector corresponding to the zero eigenvalue has a
certain number of components that are of order one, whereas all
others are very small. The number of components of normal size
determines the optimal number of active wells: only one well if
$\nu$ is larger than 10, four if $\nu=3$, etc. The eigenvector
of the reduced system of equations for the active wells is a
good approximation to the large components of the eigenvector
corresponding to the complete system. As we approach the
continuum limit, more and more points enter the reduced system
of equations and the approach of Section \ref{sec:3} becomes a
viable alternative to these methods.

As before, the wavefront profile will obey $E(i-ct)=E_i(t)$,
where $i=-L,\ldots,M$ are the indices of the $A=L+M+1$ active
wells. Then their reduced dynamics obeys (\ref{m8}) for
$i=-L,\ldots,M$ together with the approximations $E_{-L-1}\sim
E^{(1)}$ and $E_{M+1}\sim E^{(3)}$. We have to find approximate
stationary solutions of this system, and then linearize it with
respect to the appropriate one. At the approximate critical
current $J_c$, one of the eigenvalues of the linearized system
becomes zero. Let us call $U_i^\dag$ and $U_i$ the
$A$-dimensional ($A=L+M+1$) left and right eigenvectors
corresponding to the zero eigenvalue of the coefficient matrix
for the linearized system, chosen so that $\sum_{i=-L}^M
U_i^\dag U_i =1$. Very close to $J_c$, the electric field
profile will be $E_i(t)\sim E_i(J_c,\nu) + U_i \varphi(t)$ plus
terms which decrease exponentially fast as time elapses.
$\varphi$ obeys Eq. (\ref{m12}) with the following coefficients
\begin{eqnarray}
\alpha = \sum_{i=-L}^M U_i^\dag + {v_{0}+D_{0}\over\nu\,
v'_{-L-1}}\, U_0^\dag + {D_{M}U_{M}^{\dag}\over\nu\,
v'_{M+1}}\,,\label{m15}\\
\beta = {1\over 2\nu}\, \sum_{i=-L}^M \left[D''_i  U_i^2
(E_{i+1} + E_{i-1} -2 E_{i}) \right.\nonumber\\
+2 D'_i U_i (U_{i+1}+U_{i-1} - 2U_i) - 2 v'_{i} U_i (U_i -
U_{i-1})
\nonumber\\
\left. + v''_0 (E_{i-1} - E_{i}- 2\nu)U_i^2 \right]\,
U_i^\dag . \label{m16}
\end{eqnarray}
Here we have set $U_{-L-1}=U_{M+1}=0$, $E_{-L-1}= E^{(1)}$,
$E_{M+1}=E^{(3)}$, $v_i= v(E_i)$, $v'_i= v'(E_i)$ $D_i=D(E_i)$,
etc. All quantities are evaluated at $J=J_c$. Clearly these
formulas become (\ref{m13}) and (\ref{m14}) if $L=M=0$, $U_0=
U_0^\dag = 1$.

Obviously, the solution and interpretation of (\ref{m12}) is
the same as before, and we only have to change the matching
condition and the description of the jump of the active wells.
Increasing the number of active wells improves appreciably the
approximation of the wavefront velocity; see Fig.\ \ref{fig5}.
The jump of the active wells is described by the solution of
(\ref{m8}) for $i=-L,\ldots,M$, with $J=J_c$, $E_i(t)= E_i(J_c,
\nu) + U_i \varphi(t)$, and the same matching condition as
before. Fig.\ \ref{fig8} depicts a comparison of the composite
approximation and the numerically calculated field profile for
$\nu=3$. Notice the improvement with respect to the single
active well approximation.

Our analysis shows that the transition from moving to stationary
fronts involves loss of continuity of the wavefront profile.
For the reduced system of active wells, the transition is a
global saddle-node bifurcation: two stationary solutions
coalesce at the critical current and (for $J<J_1(\nu)$ or
$J>J_2(\nu)$), a traveling wavefront appears. Near the critical
currents, the field profile of this front is very sharp: it
resembles the typical discrete steps of the stationary profile
on long intervals of time (scaling as $|J-J_i|^{-{1\over 2}}$),
followed by abrupt transitions between steps. Although we do
not have a proof, it is natural to conjecture that the
depinning transition for the complete SL model is of the same
type as that for the reduced system of active wells.

\noindent {\em Remark}. Notice that our theory of front
depinning applies with minor modifications to discrete models
with more complicated (than (\ref{m3})) constitutive relations,
$J_{i\to i+1} = {\cal J}(F_i,n_i, n_{i+1})$ (cf Wacker's paper
in Ref. \onlinecite{bon94}) or $J_{i\to i+1} = {\cal J}(F_{i-1},
F_{i},F_{i+1},n_i, n_{i+1})$ (cf the complete discrete model in
Ref. \onlinecite{bon00} where potential drops at the barriers,
$V_i$, are used instead of electric fields averaged over one SL
period, $F_i$). The key point is that only a finite number of
wells are active in wavefront depinning for sufficiently large
$\nu$.

\section{Continuum limit}
\label{sec:3}
The continuum limit of the DDD model is useful to understand
self-sustained oscillations of the current and wavefront
motion. It consists of $\nu\to 0$, $i\to\infty$, with $\nu i =
x\in [0,N\nu]$, $N\nu\gg 1$. In this limit, (\ref{m8})
becomes
\begin{eqnarray}
{\partial E\over\partial t} + v(E)\, {\partial E\over\partial
x} = J - v(E), \label{c1}
\end{eqnarray}
up to terms of order $\nu$. Equation (\ref{c1}) corresponds to
the hyperbolic limit of the well-known Kroemer model of the
Gunn effect \cite{hig92}. With constant $J$, shock waves are
solutions of these equations \cite{kni66}. These waves are
related to wavefronts, and their speed can be calculated
explicitly. Let $V(E_+,E_-)$ denote the speed of a shock wave
such that $E$ becomes $E_-$ (resp.\ $E_+$) to the left (resp.\
right) of the shock. Inside the shock wave, we should use the
discrete model with $n_i\gg 1$, $dE_i/dt \sim -V\, (E_i-
E_{i-1})/\nu\gg 1$, and $J=O(1)$. Notice that we have rescaled
the wavefront velocity so that $V = c\nu$ is the correct
velocity for the (rescaled) continuum profile $E(x-Vt) = E(\nu\,
(i - ct))$. Then
\begin{eqnarray}
E_+ - E_- = \sum (E_i-E_{i-1}) \sim \nu \sum n_i\nonumber\\
\sim V \sum {E_{i}-E_{i-1}\over v(E_{i})+D(E_{i})} + \sum
{D(E_{i})\over v(E_{i}) + D(E_{i})}\nu n_{i+1}\nonumber\\
\sim V \sum {E_{i}-E_{i-1}\over v(E_{i})+D(E_{i})} + \sum
{D(E_{i})\, (E_{i+1}-E_{i})\over v(E_{i}) + D(E_{i})}\,.
\nonumber
\end{eqnarray}
This expression yields
\begin{eqnarray}
V\sim {\sum {v(E_{i})\, (E_{i+1}-E_{i})\over v(E_{i}) +
D(E_{i})}\over \sum {E_{i}-E_{i-1}\over v(E_{i})+D(E_{i})}} \,,
\nonumber
\end{eqnarray}
which becomes
\begin{eqnarray}
V(E_+,E_-) = {\int_{E_{-}}^{E_{+}} {v(E)\over v(E) + D(E)}\,
dE\over \int_{E_{-}}^{E_{+}} {dE\over v(E)+D(E)}}\,, \label{c4}
\end{eqnarray}
in the continuum limit. This expression is equivalent to the
following weighted equal-area-rule
\begin{eqnarray}
\int_{E_{-}}^{E_{+}} {v(E)-V(E_{+},E_{-})\over v(E) + D(E)}\,
dE = 0. \label{c5}
\end{eqnarray}
For $D=0$, this formula reduces to that derived for the
discrete drift model in Ref.\ \onlinecite{bon95}. An expression
including the next correction to this formula is given in
Appendix \ref{ear}. There is only one value of $J$, $J^*$, such
that $V=J$ with $E_- = E^{(1)}(J)$ and $E_+ = E^{(3)}(J)$. For
$J\in (v_m,J^*)$, a wavefront joining $E^{(1)}(J)$ to
$E^{(3)}(J)$ consists of a shock wave having $E_+= E^{(3)}(J)$,
and $E_-$ such that $V(E^{(3)}(J),E_-) =v(E_-)$. Furthermore,
to the left of the shock wave, there is a {\em tail} region
moving rigidly with the shock and such that
\begin{eqnarray}
[v(E)-V]\, {\partial E\over\partial\xi} = J  - v(E),
\label{c6}
\end{eqnarray}
for negative $\xi = x-Vt$, and $E(-\infty)=E^{(1)}(J)$,
$E(0)=E_-$. This whole structure (shock and tail region) is
called a {\em monopole with left tail} \cite{bon97}. Similarly,
for $J\in (J^*,1)$, a wavefront joining $E^{(1)}(J)$ to
$E^{(3)}(J)$ becomes a {\em monopole with right tail}. This
monopole consists of a shock wave having $E_-= E^{(1)}(J)$, and
$E_+$ such that $V(E_+,E^{(1)}(J)) =v(E_+)$, and a tail region
satisfying (\ref{c6}) for positive $\xi$, with the boundary
conditions $E(0)=E_+$ and $E(\infty)=E^{(3)}(J)$ \cite{bon97}.
In conclusion, the wavefront velocity as a function of $J$ is
determined by the following equations:
\begin{eqnarray}
C(J) &=& V(E^{(3)}(J),E_-),\,\mbox{with}\nonumber\\
v(E_-) &=& V(E^{(3)}(J),E_-),\,\mbox{if}\quad v_m<J<J^*,
\label{c7}\\
C(J) & =& V(E_+,E^{(1)}(J)),\,\mbox{with}\nonumber\\
v(E_+) &=& V(E_+,E^{(1)}(J)),\,\mbox{if}\quad J^*<J<1.
\label{c8}
\end{eqnarray}
Notice that this $C(J)$ is the limiting value of the rescaled
wavefront velocity, $c(J,\nu)\nu$ as $\nu\to 0+$. We have
compared the continuum approximation of the wavefront velocity
(in wells traversed per unit time, i.e., $c(J,\nu)=C(J)/\nu$,
not rescaled) with the numerical solution of the model for
$\nu=0.01$ in Fig.\ \ref{fig9}. The equal-area rule result
corresponds to (\ref{c4}), (\ref{c7}) and (\ref{c8}) and its
maximum difference with the numerical solution is about 17.6 \%.
Notice that the corrections in Appendix \ref{ear} improve
significantly the result: the corrected equal-area result
(\ref{a4}) differs at most 3\% from the numerical solution.
\\

\noindent {\em Remark}. A different strategy to derive a
formula for the shock velocity $V(E_+,E_-)$ could be to retain
the first order corrections to (\ref{c1}) and then use
well-known procedures for partial differential equations. The
first-order correction to (\ref{c1}) is
\begin{eqnarray}
{\partial E\over\partial t} + v(E)\, {\partial E\over\partial
x} = J - v(E) \nonumber\\
+ \nu\,\left(D(E) + {v(E)\over 2}\right) {\partial^{2}
E\over\partial x^{2}}\,,\label{c9}
\end{eqnarray}
in which terms of order $\nu^2$ are ignored. As $\nu\to 0$,
this equation has shock waves whose velocity obeys the
equal-area rule
\begin{eqnarray}
\int_{E_{-}}^{E_{+}} {v(E)-V(E_{+},E_{-})\over D(E) +
{v(E)\over 2} }\, dE = 0. \label{c10}
\end{eqnarray}
instead of (\ref{c5}). Notice that this rule gives the same
result as (\ref{c5}) for $D=0$. What is wrong in this argument?
Notice that we have to use the rescaled moving variable $\xi=
(x-Vt)/\nu$ to derive (\ref{c10}). After rescaling, both
convective and diffusive terms in (\ref{c9}) are the largest
ones, of order $\nu^{-1}\gg 1$. But so are all the terms in the
Taylor series $(E_i -E_{i-1})/\nu = \partial E/\partial x -
(\nu/2)\,\partial^{2} E/\partial x^{2} + (\nu^2/6)\,
\partial^{3} E/\partial x^{3} + \ldots$, which was used to
derive (\ref{c9}) in the first place. Thus this derivation
ignores infinitely many relevant terms in the Taylor expansions
of $n_i$ and $n_{i+1}$ and it yields the incorrect formula
(\ref{c10}) as a result.

\section{Conclusions}
\label{sec:conclusions}
We have analyzed the motion of wavefronts in discrete
drift-diffusion models of nonlinear charge transport in
superlattices. Moving wavefronts are profiles of the electric
field traveling rigidly with constant velocity. Propagation
failure of the fronts occurs because the field profile loses
continuity at the critical currents and becomes pinned at
discrete sites. We have characterized propagation failure of
wavefronts and, conversely, front depinning by singular
perturbation methods. These methods are based upon the fact
that only a few active wells characterize wavefront motion for
dimensionless doping $\nu$ sufficiently large. In the continuum
limit, as $\nu$ tends to zero, more and more wells become
active and a different approximation makes sense. In this
limit, discrete equations turn into differential equations, and
wavefronts turn into monopoles which are shock waves with a
rigid tail region to their left or right. We have derived the
equal area rule for such shock waves, and also its leading
correction. Our different asymptotic theories perform well in
their respective domains of validity and approximate the motion
of a wavefront at constant current on an infinite superlattice.
Understanding this motion is the key to understanding more
sophisticated phenomena occuring in semiconductor superlattices
under voltage bias conditions\cite{kas95,bon97}.

\acknowledgements
This work has been supported by the Spanish DGES through grant
PB98-0142-C04-01.

\appendix
\setcounter{equation}{0}
\section{Drift velocity and diffusivity coefficients}
\label{app1}
To compare our theoretical results with those of numerical
solutions of the model, it is better to use analytical
approximations of drift velocity and diffusivity instead of
using numerical data for them. We use the following
dimensionless coefficients which approximately fit those used
in Ref.\ \onlinecite{car00}:
\begin{eqnarray}
v(E) &=& {0.2684\, E\over 0.25 + (0.9862 E - 0.85)^{2}}
\nonumber\\
&+& {0.9072\over 0.16 + (0.9862 E
-15)^{2} } - 0.004 , \label{a1}\\
D(E) &=& 2\, e^{-E^{2}}\,.\label{a2}
\end{eqnarray}
The velocity has a local maximum at $E=1$, $v=1$.

\setcounter{equation}{0}
\section{First order correction to equal-area rule}
\label{ear}
A more careful derivation of the equal-area rule uses the
trapezoid rule for Riemann integrals:
\begin{eqnarray}
\int_{E_{-}}^{E_{+}} f(E) dE &\approx & \sum f(E_i)\, (E_i -
E_{i-1})
\nonumber\\
&-& {1\over 2}\, \sum [f(E_i) - f(E_{i-1})](E_i -
E_{i-1})\nonumber\\
&\approx & \sum f(E_{i-1})\, (E_i - E_{i-1})
\nonumber\\ &+& {1\over 2}\, \sum [f(E_i) -
f(E_{i-1})] (E_i - E_{i-1}) .\label{a3}
\end{eqnarray}
We can repeat our derivation in Section \ref{sec:3} step by
step, but keeping now the correction terms (\ref{a3}) when
substituting integrals instead of Riemann sums. The result is
\begin{eqnarray}
V(E_+,E_-) = {\int_{E_{-}}^{E_{+}} {v(E)\over v(E) + D(E)}\,
dE + S_{N}\over \int_{E_{-}}^{E_{+}} {dE\over v(E)+D(E)}+ S_{D}
}\,, \label{a4}\\
S_N = {1\over 2}\sum\left({D_{i}\over v_{i} +D_{i}} - {D_{i-1}
\over v_{i-1} +D_{i-1}}\right) (E_{i}-E_{i-1})\nonumber\\
= -{1\over 2}\sum\left({v_{i}\over v_{i} +D_{i}} - {v_{i-1}
\over v_{i-1} +D_{i-1}}\right) (E_{i}-E_{i-1})\,, \label{a5}\\
S_D = {1\over 2}\sum\left({1\over v_{i} +D_{i}} - {1\over
v_{i-1} +D_{i-1}}\right) (E_{i}-E_{i-1})\,. \label{a6}
\end{eqnarray}
In the numerator of (\ref{a4}), we have ignored the term
$\nu\sum (v_i - J)/(v_i + D_i)$, which vanishes in the
continuum limit. Notice that $v_{i-1}\sim v_i - v'_i\, (E_i -
E_{i-1})$, a similar relation for $D_{i-1}$ and that $v'_{i}
\leq 0$ (typically) and $D'_{i}\leq 0$ suggest that $S_N \leq 0$
and $S_D\geq 0$. Thus we expect that the corrected shock
velocity be {\em lower} than the leading order (\ref{c4}).

\begin{figure}
\includegraphics[width=8cm]{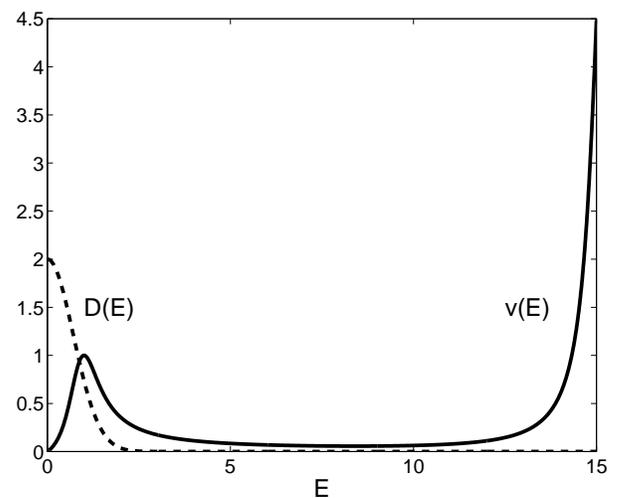}
\vspace{0.5 cm}
\caption{Electron drift velocity $v(E)$ and diffusion
coefficient $D(E)$ as functions of the electric field in
nondimensional units. }
\label{fig2}
\end{figure}

\begin{figure}
\includegraphics[width=8cm]{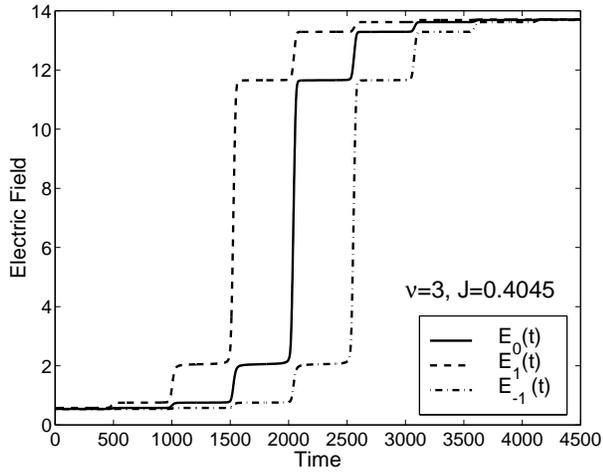}
\vspace{0.5 cm}
\caption{Electric field profile of a wavefront at three
consecutive wells. We have depicted the field at each well as a
function of time, which illustrates the motion of the front at
constant velocity. }
\label{fig3}
\end{figure}

\begin{figure}
\includegraphics[width=8cm]{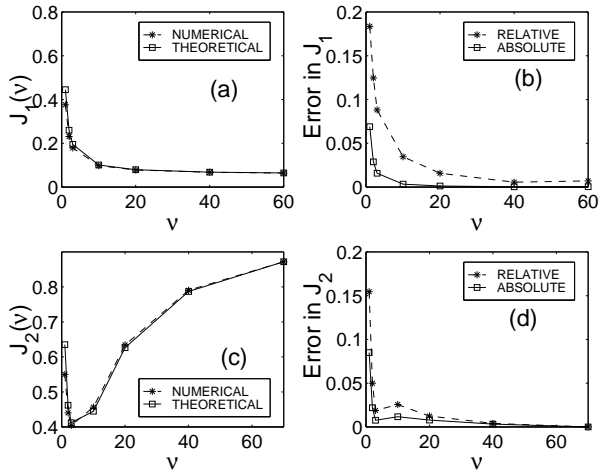}
\vspace{0.5 cm}
\caption{Upper and lower critical currents as functions of the
dimensionless $\nu$. }
\label{fig4}
\end{figure}

\begin{figure}
\includegraphics[width=8cm]{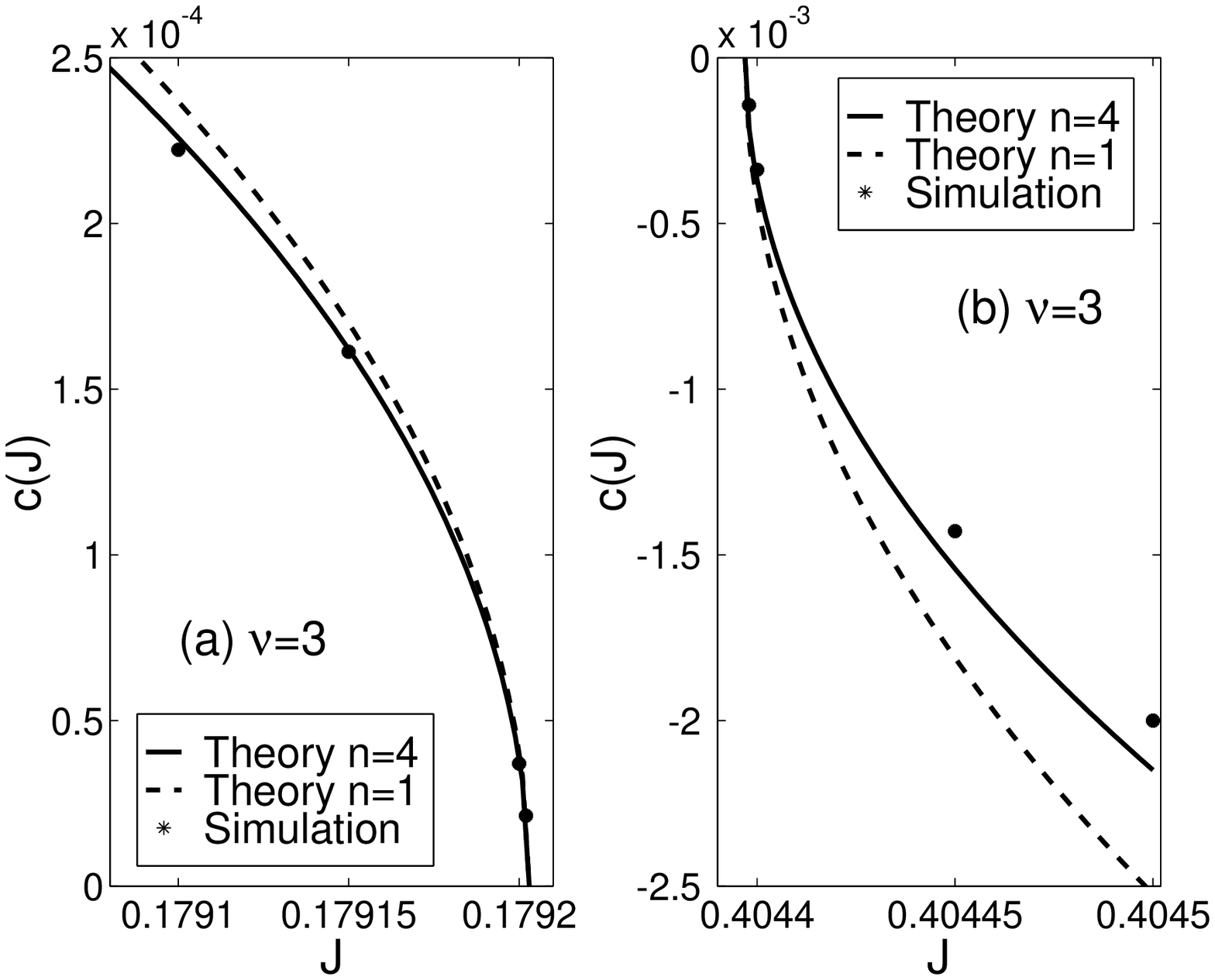}
\vspace{0.5 cm}
\caption{Wavefront velocity as a function of current density
for $\nu=3$. We have compared the numerically measured velocity
to the results of our theory with one or four active wells. }
\label{fig5}
\end{figure}

\begin{figure}
\includegraphics[width=8cm]{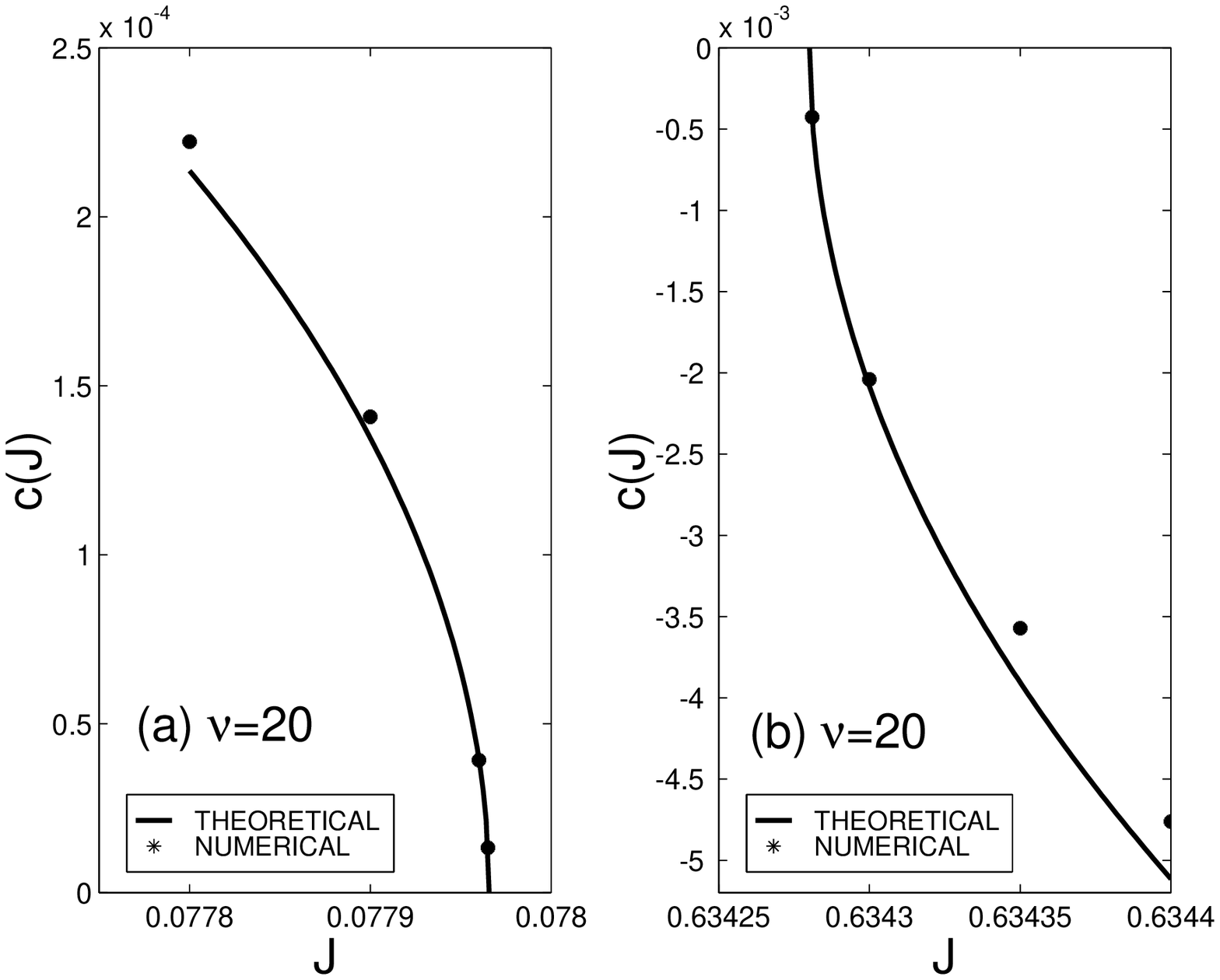}
\vspace{0.5 cm}
\caption{Wavefront velocity as a function of current density
for $\nu=20$. We have compared the numerically measured velocity
to the results of our theory with one active well.}
\label{fig6}
\end{figure}

\begin{figure}
\begin{center}
\includegraphics[width=8cm]{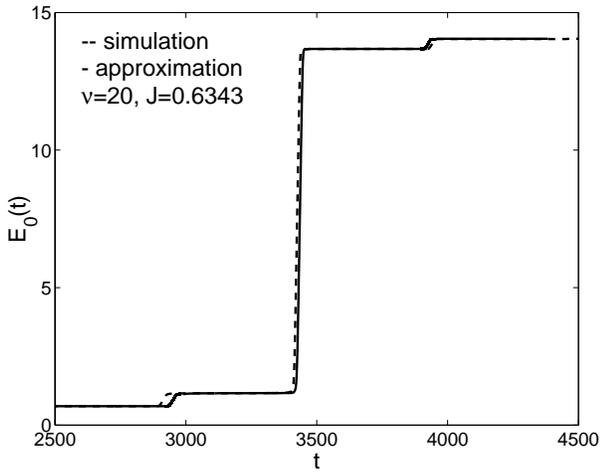}
\caption{Wavefront profiles near $J=J_2$ for $\nu=20$. We
compare the results of matched asymptotic expansions with one
active well and the numerical solution of the model.  }
\label{fig7}
\end{center}
\end{figure}

\begin{figure}
\begin{center}
\includegraphics[width=8cm]{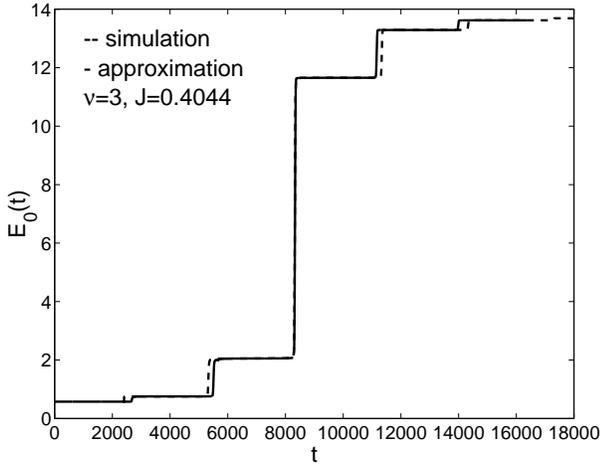}
\caption{Wavefront profiles near $J=J_2$ for $\nu=3$. We
compare the numerical solution of the model with the result of
matched asymptotic expansions with 5 active wells. Notice that
the largest error source is the theoretical estimate of the
time period (an error of 200 in a 3,000 period). }
\label{fig8}
\end{center}
\end{figure}

\begin{figure}
\begin{center}
\includegraphics[width=8cm]{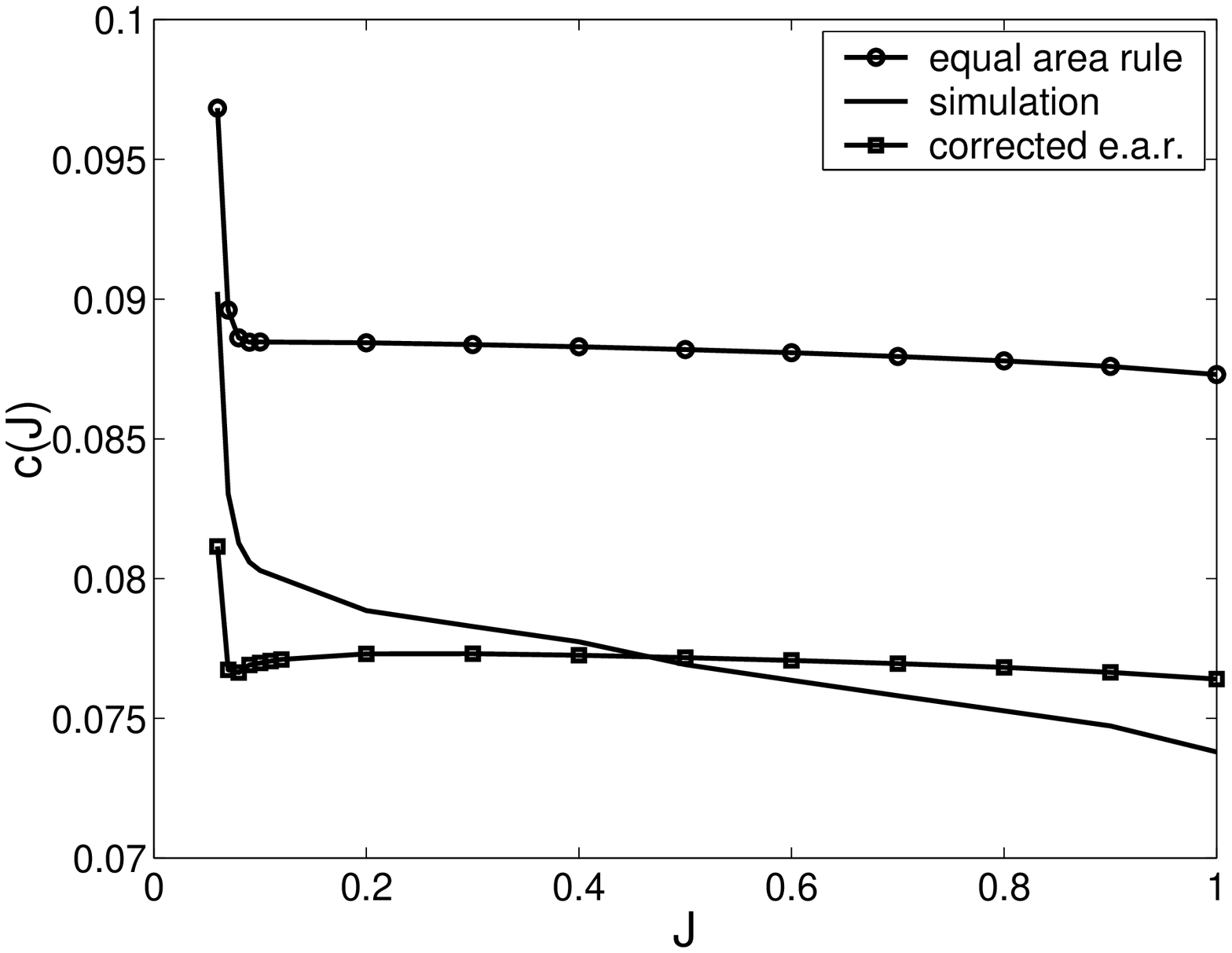}
\caption{Comparison of the equal-area rule (leading order) and
corrected equal-area rule (including first order corrections)
approximations to the wavefront velocity with the numerical
solution of the model for $\nu=0.01$. }
\label{fig9}
\end{center}
\end{figure}

\end{multicols}

\begin{references}{}
\bibitem{gunn}
Refs.\ \onlinecite{bon97} and \onlinecite{hig92} consider
current self-oscillations mediated by monopoles in SL with
$D=0$ or by dipoles in the hiperbolic limit of the Gunn effect,
respectively. We can use these studies to describe
self-oscillations in SL with $D\neq 0$. The only thing we need
to change is to substitute our equal-area formulas (\ref{c4}) or
(\ref{a4}) instead of the equal-area formulas considered in
those papers.
\end{references}
\end{document}